\documentstyle[12pt,amsfonts]{article}
\newcommand{\svec}[1]{ \stackrel{\rightharpoonup}{#1} }
\newcommand{\svvec}[1]{ \stackrel{\rightharpoonup \! \! \! \rightharpoonup}{#1} }

\newcommand{\tdot}[1]{ \stackrel{\cdot}{#1} }

\newcommand{\define}{ \stackrel{\triangle}{=} }

\def\be{\begin{equation}}
\def\ee{\end{equation}}
\def\ba{\begin{array}}
\def\ea{\end{array}}

\def\d4{{\rm d}^4}

\begin{document}
\title{\bf Mechanism of Gravity Impulse}
\author{ {Ning WU}\thanks{email address: wuning@mail.ihep.ac.cn}
\\
\\
{\small Institute of High Energy Physics, P.O.Box 918-1,
Beijing 100039, P.R.China}
}
\maketitle


\begin{abstract}
It is well-known that energy-momentum is the source of gravitational field.
For a long time, it is generally believed that only stars with huge
masses can generate strong gravitational field. Based on the
unified theory of gravitational interactions and electromagnetic
interactions, a new mechanism of the generation of gravitational
field is studied. According to this mechanism, in some special
conditions, electromagnetic energy  can be directly converted
into gravitational energy, and strong gravitational field can be
generated without massive stars. Gravity impulse found in
experiments is generated by this mechanism.
\\

\end{abstract}

~~


\newpage

\Roman{section}

\section{Introduction}

Gravity is one kind of fundamental interactions
which is known by human beings in very ancient times. In
Newton's classical theory of gravity, gravity  obeys the
inverse square law and the magnitude of gravity is proportional
to the mass of the object\cite{n01}. In Einstein's general theory
of gravity, gravity is treated as space-time geometry\cite{n02,n03}.
In both Newton's classical theory of gravity and Einstein's
general relativity, gravitational field is generated by
energy-momentum of matter field, and there is no other way that
can directly convert energy of electromagnetic field into
energy of gravitational field. \\

Quantum Gauge Theory of Gravity(QGTG) is first proposed in
2001\cite{w01,w02,w03,w04}. It is a quantum theory of gravity
proposed in the framework of quantum gauge field theory.
An important breakthrough on QGTG
is obtained in 2003, when Quantum Gauge General Relativity(QGGR) is proposed
in the framework of quantum gauge theory of gravity\cite{w05,w06}.
In QGGR, the field equation of gravitational gauge field is just
Einstein's field equation, so in classical level, we can set up
its geometrical formulation\cite{w061}, and QGGR returns to
Einstein's general relativity in classical level. In QGGR, the equation
of motion of a mass point in gravitational field is given by Newton's
second law of motion, which is equivalent to the geodesic
equation in general relativity\cite{w07,w08}. For classical
tests of gravity, QGGR gives out the same theoretical predictions
as those of general relativity\cite{w08}.  \\

In quantum level, QGTG is a perturbatively renormalizable
quantum theory, so based on it, quantum effects of gravity
and gravitational interactions of some basic fields\cite{w09,w10}
can be explored. In QGTG,
unifications of fundamental interactions including gravity can
be fulfilled in a simple and beautiful way\cite{w11,w12,w13}.
Gravitational phase effects found in COW experiments\cite{a01,a02,a03}
and gravitationally bound quantized states found recently\cite{a04,a05}
can be understood in a natural way\cite{w14}. Besides, QGTG
possibly solved the cosmological constant problems\cite{w15},
and predicts the gravitational Aharonov-Bohm effect\cite{w14}.
In the surface of a neutron star or near black hole, the
gravitomagnetic field is relatively strong, and measuring the
position of the spectral line of the radiation or absorption
induced by gravitomagnetic field can help us to determine
the strength of the gravitomagnetic field in the surface of
the star\cite{w16}. If we use the mass generation mechanism which
is proposed in literature \cite{w17}, we can propose a new
theory on gravity which contains massive graviton and
the introduction of massive graviton does not affect
the strict local gravitational gauge symmetry of the
theory and does not affect the traditional long-range
gravitational force\cite{w18}. The existence of massive graviton
will help us to understand the possible origin of dark
energy and dark matter in the Universe. \\

There is another important quantum effect of gravitational
interactions, the gravitational shielding effect found by
Podkletnov\cite{a06,a07}. This effect originates from the
interactions between gravitational field and non-homogeneous
vacuum, which make gravitational field obtain a small
mass term\cite{w19}. In 2001, Podkletnov found that discharges
originating from a superconducting ceramic electrode are
accompanied by the emission of gravitational impulse\cite{a08,a09}.
Gravity impulse can not be understood in Einstein's general
relativity. For a long time, its mechanism is unknown. Some
people  think that the phenomenon of gravity impulse is not
consistent with Einstein's general relativity, and therefore
the results is not true. Some even doubt the reality of
the experiments. In this paper, based on the unified
theory of gravitational interactions and electromagnetic
interactions, a new mechanism of the generation of gravitational
field is studied. This mechanism can be used to explain the
gravity impulse found by E.Podkletnov. \\

\section{GU(1) Unification Theory}
\setcounter{equation}{0}

In QGTG\cite{w01,w02,w03,w04,w05,w06}, the most
fundamental quantity is gravitational gauge field $C_{\mu}(x)$,
which is the gauge potential corresponding to gravitational
gauge symmetry. Gauge field $C_{\mu}(x)$ is a vector in
the gravitational Lie algebra, which can be expanded as
\be \label{2.1}
C_{\mu}(x) = C_{\mu}^{\alpha} (x) \hat{P}_{\alpha},
~~~~~~(\mu, \alpha = 0,1,2,3)
\ee
where $C_{\mu}^{\alpha}(x)$ is the component field and
$\hat{P}_{\alpha} = -i \frac{\partial}{\partial x^{\alpha}}$
is the  generator of the global gravitational
gauge group. The gravitational gauge covariant derivative is given by
\be \label{2.2}
D_{\mu} = \partial_{\mu} - i g C_{\mu} (x)
= G_{\mu}^{\alpha} \partial_{\alpha},
\ee
where $g$ is the gravitational coupling constant and matrix
$G$ is defined by
\be \label{2.3}
G = (G_{\mu}^{\alpha}) = ( \delta_{\mu}^{\alpha} - g C_{\mu}^{\alpha} ).
\ee
Its inverse matrix is denoted as $G^{-1}$
\be \label{2.4}
G^{-1} = \frac{1}{I - gC} = (G^{-1 \mu}_{\alpha}).
\ee
Using matrix $G$, $G^{-1}$ and Minkowski metric $\eta$, we can
define two important composite operators
\be \label{2.5}
g^{\alpha \beta} = \eta^{\mu \nu}
G^{\alpha}_{\mu} G^{\beta}_{\nu},
\ee
\be \label{2.6}
g_{\alpha \beta} = \eta_{\mu \nu}
G_{\alpha}^{-1 \mu} G_{\beta}^{-1 \nu},
\ee
which are widely used in QGTG. In QGTG,
 space-time is always flat and space-time metric
is always the Minkowski metric, so $g^{\alpha\beta}$ and $g_{\alpha\beta}$
are no longer space-time metric. They are only two composite operators
which  consist of gravitational gauge field.
\\

The  field strength of gravitational gauge field is defined by
\be \label{2.7}
F_{\mu\nu} \define \frac{1}{-ig} \lbrack D_{\mu}~~,~~D_{\nu} \rbrack.
\ee
Its explicit expression is
\be \label{2.8}
F_{\mu\nu}(x) = \partial_{\mu} C_{\nu} (x)
-\partial_{\nu} C_{\mu} (x)
- i g C_{\mu} (x) C_{\nu}(x)
+ i g C_{\nu} (x) C_{\mu}(x).
\ee
$F_{\mu\nu}$ is also a vector in gravitational Lie algebra,
\be \label{2.9}
F_{\mu\nu} (x) = F_{\mu\nu}^{\alpha}(x) \cdot \hat{P}_{\alpha},
\ee
where
\be \label{2.10}
F_{\mu\nu}^{\alpha} = \partial_{\mu} C_{\nu}^{\alpha}
-\partial_{\nu} C_{\mu}^{\alpha}
-  g C_{\mu}^{\beta} \partial_{\beta} C_{\nu}^{\alpha}
+  g C_{\nu}^{\beta} \partial_{\beta} C_{\mu}^{\alpha}.
\ee
Using matrix $G$, its expression can be written in a
simpler form
\be \label{2.11}
F_{\mu\nu}^{\alpha} =
G_{\mu}^{\beta} \partial_{\beta} C_{\nu}^{\alpha}
-G_{\nu}^{\beta} \partial_{\beta} C_{\mu}^{\alpha}.
\ee
\\

In the unified theory of gravitational interactions and electromagnetic
interactions, the lagrangian of theory is selected to be\cite{w11,w06}:
\be \label{2.12}
\ba{rcl}
{\cal L}_0 & = & - \bar{\psi}
\lbrack \gamma^{\mu} ( D_{\mu} -ie A_{\mu} ) + m \rbrack \psi
- \frac{1}{4} \eta^{\mu \rho} \eta^{\nu \sigma}
{\mathbb A}_{\mu \nu} {\mathbb A}_{\rho \sigma} \\
&&\\
&& - \frac{1}{16} \eta^{\mu \rho} \eta^{\nu \sigma}
g_{ \alpha \beta} F^{\alpha}_{\mu\nu}
F^{\beta}_{\rho\sigma}
- \frac{1}{8} \eta^{\mu \rho} G^{-1 \nu}_{ \beta}
G^{-1 \sigma}_{ \alpha } F^{\alpha}_{\mu\nu}
F^{\beta}_{\rho\sigma}
+ \frac{1}{4} \eta^{\mu \rho} G^{-1 \nu}_{ \alpha}
G^{-1 \sigma}_{ \beta } F^{\alpha}_{\mu\nu}
F^{\beta}_{\rho\sigma},
\ea
\ee
where  $A_{\mu}$ is the electromagnetic field, and
\be \label{2.13}
{\mathbb A}_{\mu\nu} = D_{\mu} A_{\nu} - D_{\nu} A_{\mu}
+ g G^{-1 \lambda}_{\alpha} A_{\lambda} F_{\mu\nu}^{\alpha}
\ee
is the gravitational gauge covariant field strength of electromagnetic
field. This system has $GU(1)$ gauge symmetry\cite{w11}. \\

\section{Field Equation of Electromagnetic Field in Gravitational Field}
\setcounter{equation}{0}

Euler-Lagrangian equation gives out the following field equation of
electromagnetic field
\be \label{3.1}
\ba{rcl}
\eta^{\mu\rho} (D_{\mu} {\mathbb A}_{\rho\nu})
&=&  - J^{\mu} \eta_{\mu\nu}
- \eta^{\lambda\rho} (\partial_{\mu} G_{\lambda}^{\mu})
{\mathbb A}_{\rho\nu}  \\
&&\\
&& + \frac{g}{2} \eta^{\mu\rho} \eta^{\lambda\sigma} \eta_{\nu\nu_1}
G^{-1 \nu_1}_{\alpha} {\mathbb A}_{\rho\sigma} F_{\mu\lambda}^{\alpha}
- g \eta^{\lambda\rho} G^{-1 \kappa}_{\alpha} (D_{\lambda} C_{\kappa}^{\alpha})
{\mathbb A}_{\rho\nu},
\ea
\ee
where  $J^{\mu}$  is the electric current of Dirac field
\be \label{3.2}
J^{\mu} = i e \bar{\psi} \gamma^{\mu} \psi.
\ee
The generalized Bianchi identity for electromagnetic field is
\be \label{3.4}
D_{\mu} A_{\nu\lambda} +  D_{\nu} A_{\lambda\mu}
+ D_{\lambda} A_{\mu\nu} =
- F^{\alpha}_{\mu\nu} (\partial_{\alpha} A_{\lambda})
- F^{\alpha}_{\nu\lambda} (\partial_{\alpha} A_{\mu})
- F^{\alpha}_{\lambda\mu} (\partial_{\alpha} A_{\nu}).
\ee
\\

The above two equations (\ref{3.1}) and (\ref{3.4}) are
strict relations. Now, let's consider their approximations
in the limit of weak gravitational gauge field. Supposed that
gravitational field $gC_{\mu}^{\alpha}$ is a first order
infinitesimal quantity, then in first order approximation,
equation (\ref{3.1}) gives out the following two equations
\be \label{3.5}
\nabla \cdot \svec{E}_e = - J_0
+ g \partial_{\lambda} (C_{i}^{\lambda} A_{i0})
- \frac{g}{2} \eta^{\mu\rho} \eta^{\lambda\sigma}
A_{\rho\sigma} F_{\mu\lambda}^0
- g (\partial_i C^{\alpha}_{\alpha}) A_{i0}
- g \partial_i (A_{\alpha} F_{i0}^{\alpha}),
\ee
\be \label{3.6}
\ba{rcl}
\left ( \frac{\partial}{\partial t} \svec{E}_e
- \nabla \times \svec{B}_e \right )_i
& = & -  J_i
+ g \eta^{\mu\rho} \partial_{\lambda}
(C_{\mu}^{\lambda} A_{\rho i})
+ \frac{g}{2} \eta^{\mu\rho} \eta^{\lambda\sigma}
A_{\rho\sigma} F_{\mu\lambda}^i \\
&&\\
&& - g \eta^{\lambda\rho} (\partial_{\lambda} C^{\alpha}_{\alpha})
A_{\rho i}
- g \partial^{\mu} (A_{\alpha} F_{\mu i }^{\alpha}),
\ea
\ee
and the Bianchi identity (\ref{3.4})
gives out the following two equations
\be \label{3.7}
\frac{\partial}{\partial t} \svec{B}_e
+ \nabla \times \svec{E}_e
=   g C_{0}^{\alpha} (\partial_{\alpha} \svec{B}_e)
- g \svec{C}^{\alpha} \times \partial_{\alpha} \svec{E}_e
- \svec{E}^{\alpha} \times \partial_{\alpha} \svec{A}
+ (\partial_{\alpha} A_0) \svec{B}^{\alpha},
\ee
\be \label{3.8}
\nabla \cdot \svec{B}_e =
 g \svec{C}^{\alpha} \cdot (\partial_{\alpha} \svec{B}_e)
 + \svec{B}^{\alpha} \cdot (\partial_{\alpha} \svec{A}),
\ee
where
\be \label{3.9}
E_i^{\alpha} = F_{0i}^{\alpha},~~~
\svec{E}^{\alpha} = (E_1^{\alpha}, E_2^{\alpha},  E_3^{\alpha}),
\ee
\be \label{3.10}
B_i^{\alpha} = - \frac{1}{2} \varepsilon_{ijk} F_{jk}^{\alpha},~~~
\svec{B}^{\alpha} = (B_1^{\alpha}, B_2^{\alpha},  B_3^{\alpha}),
\ee
\be \label{3.11}
E_{ei} = A_{i0},~~~
\svec{E}_e = (E_{e1}, E_{e2},  E_{e3}),
\ee
\be \label{3.12}
B_{ei} =  \frac{1}{2} \varepsilon_{ijk} A_{jk},~~~
\svec{B}_e = (B_{e1}, B_{e2},  B_{e3}),
\ee
\be \label{3.13}
A_{\mu\nu} = D_{\mu} A_{\nu} - D_{\nu} A_{\mu},
\ee
\be \label{3.14}
\svec{C}^{\alpha} = (C_1^{\alpha}, C_2^{\alpha}, C_3^{\alpha}).
\ee
$\svec{E}^{\alpha}$ and $\svec{B}^{\alpha}$ are the gravitoelectric field
and gravitomagnetic field respectively, and $\svec{E}_e$ and $\svec{B}_e$
are the electromagnetic electric field and magnetic field respectively.
If gravitational field vanishes, the above equations (\ref{3.5}) -- (\ref{3.8})
return to the Maxwell equations. Equations (\ref{3.5}) and (\ref{3.6})
can be further changed into
\be \label{3.15}
\ba{rcl}
\nabla \cdot \svec{E}_e & = & - J_0
+ g \partial_{\lambda} (\svec{C}^{\lambda} \cdot \svec{E}_e)
+ g \svec{E}_e \cdot \svec{E}^0
+ g \svec{B}_e \cdot \svec{B}^0 \\
&&\\
&& -g \svec{E}_e \cdot \nabla C_{\alpha}^{\alpha}
+ g \nabla \cdot (A_{\alpha} \svec{E}^{\alpha}),
\ea
\ee
\be \label{3.16}
\ba{rcl}
\frac{\partial}{\partial t} \svec{E}_e
- \nabla \times \svec{B}_e
& = & - \svec{J}
+ g  \partial_{\lambda} (C_0^{\lambda} \svec{E}_e )
+ g \partial_{\lambda} (\svec{C}^{\lambda} \times \svec{B}_e)
- g \svec{E}_e \cdot \svvec{E}
- g \svec{B}_e \cdot \svvec{B} \\
&&\\
&& - g \svec{E}_e (\partial_0 C^{\alpha}_{\alpha})
- g \svec{B}_e \times \nabla C_{\alpha}^{\alpha}
+ g \partial_0 (A_{\alpha} \svec{E}^{\alpha})
-g \nabla \times (A_{\alpha} \svec{B}^{\alpha}),
\ea
\ee
where
\be \label{3.17}
\svvec{E} = (\svec{E}^1, \svec{E}^2, \svec{E}^3 ), ~~~
\svvec{B} = (\svec{B}^1, \svec{B}^2, \svec{B}^3 ),
\ee
\be \label{3.18}
(\svec{E}_e \cdot \svvec{E} )_i = \svec{E}_e \cdot \svec{E}^i,
\ee
\be \label{3.19}
(\svec{B}_e \cdot \svvec{B} )_i = \svec{B}_e \cdot \svec{B}^i.
\ee

\section{Mechanism of Gravity Impulse}
\setcounter{equation}{0}

It is known that superconductor is a perfect diamagnet. Inside
the superconductor, electromagnetic electric field and magnetic
field vanish, but in most general case, the electromagnetic
gauge potential $A_{\mu}$ does not vanish. Because, inside the
superconductor, electromagnetic electric field $\svec{E}_e$
vanish, in leading order approximation, we have
\be \label{4.1}
\nabla \phi = - \tdot{\svec{A}},
\ee
where
\be \label{4.2}
\phi = - A_0
\ee
is the scalar potential. In this case, the four equations
(\ref{3.15}), (\ref{3.16}),  (\ref{3.7}) and  (\ref{3.8})
are simplified to
\be \label{4.3}
g \nabla \phi \cdot \svec{E}^0 = - \rho,
\ee
\be \label{4.4}
g \tdot{\phi} \svec{E}^0  - g \tdot{A}_i \svec{E}^i
- g \nabla \phi \times \svec{B}^0 = - \svec{J},
\ee
\be \label{4.5}
\tdot{\phi} \svec{B}^0
+ \svec{E}^0 \times \tdot{\svec{A}}
+(\partial_i \phi ) \svec{B}^i = 0,
\ee
\be \label{4.6}
\svec{B}^0 \cdot \tdot{\svec{A}} = 0,
\ee
where
\be \label{4.601}
\rho = J^0 = -J_0
\ee
is the electric charge density.
From equation (\ref{4.4}), we could see that, when there is non-vanishing
electric current inside the superconductor, the gravitoelectric field
$\svec{E}^{\alpha}$ and gravitomagnetic field $\svec{B}^0$ can not both
vanish, otherwise equation (\ref{4.4}) will be violated. \\

Selecting the following coordinate system: $x-y$ plane is in the
surface of the superconducting ceramic disk, and $z$ axis is perpendicular
to the surface of the emitter and pointing to the anode.
During discharges, gravitomagnetic field $\svec{B}^0$ vanishes.
Then equation (\ref{4.4}) is simplified to
\be \label{4.7}
g \tdot{\phi} \svec{E}^0  - g \tdot{A}_i \svec{E}^i
 = - \svec{J}.
\ee
In the above equation, the dominant term in the left hand side
is $g \tdot{\phi} \svec{E}^0$, we should have
\be \label{4.8}
g \tdot{\phi} \svec{E}^0  \approx
  - \svec{J}.
\ee
So, we have
\be \label{4.9}
g  \svec{E}^0  \approx
  - \frac{\svec{J}}{\tdot{\phi}}.
\ee
Before discharges, the electric scalar potential $\phi$ of the
emitter is negative, and after discharges, it vanishes. Therefore
during discharges, $\tdot{\phi}$ is positive
\be \label{4.10}
 \tdot{\phi} >0.
\ee
From above equation, we know that  $\svec{E}^0$ and $-\svec{J}$ have
the same direction. The direction of electric current is along the
minus $z$ axis, so $g \svec{E}^0$ is along $z$ axis. \\

$g \svec{E}^0$ is the gravitoelectric field. When a mass point
is in this gravitational field, it feels the following gravitational
force
\be \label{4.11}
\svec{f} = g m \svec{E}^0  ,
\ee
where $m$ is the mass of the mass point. And this force is just
along the $z$ direction. This just explains what observed in the
experiments: the radiation emitted during discharges exerts a short
repulsive force on small movable objects along the propagation
axis, and the impulse is proportional to the mass of the objects
and independent on their composition\cite{a08,a09}. \\

\section{Summary and Discussions}

In this paper, the mechanism of gravity impulse is studied in the
framework quantum gauge theory of gravity. This mechanism is
essentially a new mechanism of generation or absorption of
gravitational field or gravitational wave. HFGW should also
be explained by this mechanism. It can also be used to help us
to construct new kind of gravitational wave detector.
\\

According to equation (\ref{4.9}), the generated gravitational field
$g \svec{E}^0$ does not depend on the mass of the source. So, strong
gravitational field can be generated without huge mass of the object.
This mechanism is quite different from the traditional mechanism
of classical Newtonian theory of gravity or Einstein's general
relativity. \\

This new mechanism is a direct result of the unified theory of
gravitational interactions and electromagnetic interactions.
Existence of the gravity impulse is a direct evidence of the
validity of the $GU(1)$ unification theory\cite{w11}.
$GU(1)$ unification theory reveals a new mechanism to
generate gravitational field, which tells us that, in some special
conditions, electromagnetic energy can be directly converted into
gravitational energy. This mechanism can help us to
utilize and control  gravity.
\\


\begin{thebibliography}{99}

\bibitem{n01} Isaac Newton, {\it Mathematical Principles of Natural Philosophy},
    (Camgridge University Press, 1934) .
\bibitem{n02} Albert Einstein, Annalen der Phys., {\bf 49} (1916) 769 .
\bibitem{n03} Albert Einstein, Zeits. Math. und Phys. {\bf 62} (1913) 225.
\bibitem{w01} Ning WU, "Gauge Theory of Gravity", hep-th/0109145
\bibitem{w02} Ning WU, Commun. Theor. Phys. (Beijing, China)
        {\bf 38} (2002): 151-156.
\bibitem{w03} Ning WU, "Quantum Gauge Theory of Gravity",
    hep-th/0112062
\bibitem{w04} Ning WU, "Quantum Gauge Theory of Gravity", talk given
        at Meeting of the Devision of Particles and Fields
        of American Physical Society at the College of
        William \& Mery(DPF2002), May 24-28, 2002,
        Williamsburg, Virgia, USA; hep-th/0207254;
        Transparancy can be obtained from:
        http://dpf2002.velopers.net/talks\_pdf/33talk.pdf
\bibitem{w05} Ning WU, Commun. Theor. Phys. (Beijing, China)
        {\bf 42} (2004): 543-552.
\bibitem{w06} Ning WU, "Renormalizable Quantum Gauge General Relativity"
        gr-qc/0309041.
\bibitem{w061} Ning WU, Commun. Theor. Phys. (Beijing, China)
        {\bf 40} (2003): 337-340.
\bibitem{w07} Ning WU, {\it classical gravitational interactions
        and gravitational Lorentz force}, (has been accepted by
        Commun. Theor. Phys.),  gr-qc/0503039.
\bibitem{w08} Ning WU, Da-Hua Zhang, {\it Classical Solution of Field
        Equation of Gravitational Gauge Field and Classical Tests
        of Gauge Theory of Gravity}, gr-qc/0508009.
\bibitem{w09} Ning WU, Commun. Theor. Phys. (Beijing, China)
        {\bf 40} (2003): 429-434.
\bibitem{w10} Ning WU, Commun. Theor. Phys. (Beijing, China)
        {\bf 41} (2004): 381-384.
\bibitem{w11} Ning WU, Commun. Theor. Phys. (Beijing, China)
        {\bf 38} (2002): 322-326.
\bibitem{w12} Ning WU, Commun. Theor. Phys. (Beijing, China)
        {\bf 38} (2002): 455-460.
\bibitem{w13} Ning WU, Commun. Theor. Phys. (Beijing, China)
        {\bf 39} (2003): 561-568.
\bibitem{a01} R.Colella, A.W.Overhauser and S.A.Werner,
        Phys.Rev.Lett. (1975):1472-1474.
\bibitem{a02} S.A.Werner, R.Colella and  A.W.Overhauser,
        Phys.Rev.Lett. (1975): 1053-1055.
\bibitem{a03} A.W.Overhauser and R.Colella,
        Phys.Rev.Lett. (1974):1237-1239.
\bibitem{a04} V.V.Nesvizhevsky, H.G.Borner et al.
        Phys. Rev. {D 67} (2003) 102002.
\bibitem{a05} A.Westphal, Diploma thesis, Institute of Physics,
        University of Heidelberg, July 2001.
\bibitem{w14} Ning WU, {\it Non-Relativistic Limit of Dirac
            Equations in Gravitational Field and Quantum
            Effects of Gravity}, (has been accepted by
            Commun. Theor. Phys.), gr-qc/0504024.
\bibitem{w15} Ning WU, Commun. Theor. Phys. (Beijing, China)
        {\bf 40} (2003): 253-256.
\bibitem{w16} Ning WU, {\it Determination of Gravitomagnetic Field
        through GRB or X-ray Pulsars},  gr-qc/0507055.
\bibitem{w17}  Ning Wu, Commun.Theor.Phys., (Beijing, China)
        {\bf 36}(2001) 169-172;hep-th/0005072.
\bibitem{w18} Ning Wu, Commun. Theor. Phys. (Beijing, China)
        {\bf 39} (2003) 671-674;hep-th/0307003.
\bibitem{a06} E.Podkletnov and R.Nieminen, Physica {\bf C 203}
    (1992) 271.
\bibitem{a07} E. Podkletnov, {\it Weak gravitational
    shielding properties of  composite bulk $YBa_2 Cu_3 O_{7-x}$
    superconductor below 70 K under electro-magnetic field},
    report MSU-chem 95, cond-mat/9701074.
\bibitem{w19} Ning WU, Commun. Theor. Phys. (Beijing, China)
        {\bf 41}(2004) 567-572; hep-th/0307225.
\bibitem{a08} E. Podkletnov, G. Modanese, {\it Impulse Gravity
        Generator Based on Charged $Y Ba_2 Cu_3 O_{7-y}$
        Superconductor with Composite Cristal Structure},
        physics/0108005.
\bibitem{a09} E. Podkletnov, G. Modanese, {\it Investigation
        of high voltage discharges in low pressure gases through
        large ceremic superconducting electrodes}, physics/0209051.



\end{thebibliography}
\end{document}